\documentclass[12pt,titlepage]{article}
\usepackage{amsmath}
\usepackage[portuguese]{babel}

\begin{document}

\title{Oscilador harm\^{o}nico:\\
uma an\'{a}lise via s\'{e}ries de Fourier\thanks{To appear in Revista Brasileira de Ensino de F\'{\i}sica (Notas e Discuss\~oes)}\\
{\small \ (Harmonic oscillator: an analysis via Fourier series)}}
\date{}

\author{A.S. de Castro\thanks{%
E-mail: castro@pq.cnpq.br} \\
\\
Departamento de F\'{\i}sica e Qu\'{\i}mica, \\
Universidade Estadual Paulista \textquotedblleft J\'{u}lio de Mesquita
Filho\textquotedblright, \\
Guaratinguet\'{a}, SP, Brasil}
\maketitle

\begin{abstract}
O m\'{e}todo de s\'{e}ries de Fourier \'{e} usado para resolver a equa\c{c}%
\~{a}o homog\^{e}nea que governa o movimento do oscilador harm\^{o}nico.
Mostra-se que a solu\c{c}\~{a}o geral do problema pode ser encontrada com
surpreendente simplicidade para o caso do oscilador harm\^{o}nico simples.
Mostra-se tamb\'{e}m que o oscilador harm\^{o}nico amortecido \'{e} suscet%
\'{\i}vel \`{a} an\'{a}lise. \newline
\newline
\noindent Palavras-chave: oscilador harm\^{o}nico, s\'{e}ries de Fourier.%
\newline
\newline
\newline
\newline
\newline
{\small \noindent The Fourier series method is used to solve the homogeneous
equation governing the motion of the harmonic oscillator. It is shown that
the general solution to the problem can be found in a surprisingly simple
way for the case of the simple harmonic oscillator. It is also shown that
the damped harmonic oscillator is susceptible to the analysis. } \newline
\end{abstract}

\section{Introdu\c{c}\~{a}o}

O prot\'{o}tipo do oscilador harm\^{o}nico simples \'{e} o sistema
massa-mola caracterizado pela equa\c{c}\~{a}o%
\begin{equation}
\frac{d^{2}x\left( t\right) }{dt^{2}}+\omega _{0}^{2}\,x\left( t\right) =0.
\label{e}
\end{equation}%
com $\omega _{0}>0$. $\,$Esta equa\c{c}\~{a}o diferencial aparece em
diversas aplica\c{c}\~{o}es e proporciona um modelo para toda e qualquer
oscila\c{c}\~{a}o de pequena amplitude. Outrossim, serve como excelente
ferramenta pedag\'{o}gica para ilustrar com simplicidade diversas t\'{e}%
cnicas de solu\c{c}\~{a}o de equa\c{c}\~{o}es diferenciais de segunda ordem.

A s\'{e}rie de Fourier \'{e} uma s\'{e}rie de senos e cossenos usada para
representar fun\c{c}\~{o}es peri\'{o}dicas e cont\'{\i}nuas por partes, e
\'{e} uma ferramenta b\'{a}sica na busca de solu\c{c}\~{o}es de equa\c{c}%
\~{o}es diferenciais. Em geral, a busca restringe-se \`{a}s solu\c{c}\~{o}es
particulares de equa\c{c}\~{o}es n\~{a}o-homog\^{e}neas. No caso do
oscilador harm\^{o}nico, as s\'{e}ries de Fourier s\~{a}o usualmente
utilizadas apenas para a busca de solu\c{c}\~{o}es particulares do oscilador
for\c{c}ado sujeito a for\c{c}as peri\'{o}dicas (veja, e.g. \cite{but}-\cite%
{fay}). Embora tais problemas possam ser abordados sem o recurso de t\'{e}%
cnicas t\~{a}o sofisticadas, incluindo at\'{e} mesmo o caso do oscilador for%
\c{c}ado, o conhecimento de t\'{e}cnicas adicionais para a solu\c{c}\~{a}o
de um dado problema \'{e} certamente enriquecedor. Al\'{e}m do mais, a equa%
\c{c}\~{a}o do oscilador harm\^{o}nico, t\~{a}o onipresente na modelagem
matem\'{a}tica de diversos sistemas f\'{\i}sicos, pode servir como laborat%
\'{o}rio para a ilustrar de maneira simplificada a aplica\c{c}\~{a}o de s%
\'{e}ries de Fourier na resolu\c{c}\~{a}o de equa\c{c}\~{o}es diferenciais
homog\^{e}neas nas disciplinas F\'{\i}sica Matem\'{a}tica e Mec\^{a}nica Cl%
\'{a}ssica dos cursos de gradua\c{c}\~{a}o em F\'{\i}sica, tanto quanto na
disciplina Matem\'{a}tica Aplicada dos cursos de gradua\c{c}\~{a}o em Matem%
\'{a}tica.

Neste trabalho ilustramos o uso de s\'{e}ries de Fourier para resolver a equa%
\c{c}\~{a}o homog\^{e}nea do oscilador harm\^{o}nico simples. O m\'{e}todo
revela-se excepcionalmente simples mas, at\'{e} onde vai o conhecimento do
autor, n\~{a}o se encontra na literatura. N\~{a}o com tanta simplicidade
assim, mostra-se afinal que o procedimento pode ser proveitoso na obten\c{c}%
\~{a}o das solu\c{c}\~{o}es do oscilador harm\^{o}nico amortecido.

\section{S\'{e}ries de Fourier}

A s\'{e}rie de Fourier que converge uniformemente para $x\left( t\right) $
no intervalo $-T/2<t<+T/2$ \'{e} dada por \cite{but}

\begin{equation}
x\left( t\right) =\frac{a_{0}}{2}+\sum\limits_{n=1}^{\infty }\left(
a_{n}\cos n\omega t+b_{n}\mathrm{\,sen\,}n\omega t\right) ,  \label{s}
\end{equation}%
onde $\omega =2\pi /T$. Os coeficientes de Fourier s\~{a}o expressos por%
\begin{eqnarray}
a_{n} &=&\frac{2}{T}\int\limits_{-T/2}^{T/2}dt\,x(t)\,\cos n\omega t, \\
&&  \notag \\
b_{n} &=&\frac{2}{T}\int\limits_{-T/2}^{T/2}dt\,x(t)\,\mathrm{sen\,}n\omega
t,
\end{eqnarray}%
e satisfazem \`{a} desigualdade de Bessel%
\begin{equation}
\frac{a_{0}^{2}}{2}+\sum\limits_{n=1}^{\infty }\left(
a_{n}^{2}+b_{n}^{2}\right) \leq \frac{2}{T}\int\limits_{-T/2}^{T/2}dt%
\,x^{2}(t).  \label{par}
\end{equation}

\section{Oscilador harm\^{o}nico simples}

A solu\c{c}\~{a}o do oscilador harm\^{o}nico expressa pela s\'{e}rie de
Fourier \'{e} uma fun\c{c}\~{a}o peri\'{o}dica de per\'{\i}odo $T$ que
envolve uma pletora de constantes desconhecidas, tais como os coeficientes
de Fourier e at\'{e} mesmo o per\'{\i}odo de osila\c{c}\~{a}o. No entanto, a
substitui\c{c}\~{a}o de (\ref{s}) em (\ref{e}) implica que%
\begin{equation}
a_{0}=0,
\end{equation}%
\begin{equation}
\sum\limits_{n=1}^{\infty }(n^{2}\omega ^{2}-\omega _{0}^{2})\left(
a_{n}\cos n\omega t+b_{n}\mathrm{\,sen\,}n\omega t\right) =0,
\end{equation}%
e em virtude da independ\^{e}ncia linear das fun\c{c}\~{o}es $\cos n\omega t$
e $\mathrm{sen\,}n\omega t$ obtemos
\begin{equation}
n^{2}\omega ^{2}-\omega _{0}^{2}=0,\quad n\geq 1.  \label{c2}
\end{equation}%
De sorte que%
\begin{equation}
x\left( t\right) =A\cos \omega _{0}t+B\mathrm{\,sen\,}\omega _{0}t,
\label{sol1}
\end{equation}%
onde%
\begin{equation}
A=\sum\limits_{n=1}^{\infty }a_{n},\quad B=\sum\limits_{n=1}^{\infty }b_{n}.
\label{sol2}
\end{equation}%
Note que a converg\^{e}ncia das s\'{e}ries patentes em (\ref{sol2}) \'{e}
assegurada pela desigualdade de Bessel. Temos ent\~{a}o que o sistema oscila
harmonicamente com per\'{\i}odo $T=2\pi /\omega _{0}$, e que (\ref{sol1}),
com suas duas constantes de integra\c{c}\~{a}o a serem ajustadas \`{a}s condi%
\c{c}\~{o}es iniciais, representa a solu\c{c}\~{a}o geral da equa\c{c}\~{a}o
do oscilador harm\^{o}nico simples.

\section{Oscilador harm\^{o}nico amortecido}

O oscilador harm\^{o}nico amortecido, que tem como prot\'{o}tipo o sistema
massa-mola com atrito linear na velocidade, \'{e} caracterizado pela equa%
\c{c}\~{a}o%
\begin{equation}
\frac{d^{2}x\left( t\right) }{dt^{2}}+2\gamma \frac{dx\left( t\right) }{dt}%
+\omega _{0}^{2}\,x\left( t\right) =0,  \label{amo}
\end{equation}%
onde $\gamma >0$ \'{e} a constante de amortecimento. Supondo que $x\left(
t\right) $ tende a zero quando $t\rightarrow \infty $, podemos escrever%
\begin{equation}
x\left( t\right) =e^{-\gamma t}\xi \left( t\right) ,  \label{amo1}
\end{equation}%
onde, para um certo $\widetilde{\gamma }<\gamma $, $\xi \left( t\right) $ n%
\~{a}o cresce mais rapidamente do que $e^{\widetilde{\gamma }t}$ quando $%
t\rightarrow \infty $, i.e.%
\begin{equation}
\underset{t\rightarrow \infty }{\lim }|e^{-\widetilde{\gamma }t}\xi \left(
t\right) |\leq M
\end{equation}%
onde $M$ \'{e} uma constante positiva. A substitui\c{c}\~{a}o de (\ref{amo1}%
) em (\ref{amo}) permite a obten\c{c}\~{a}o de uma equa\c{c}\~{a}o
diferencial sem o termo da derivada primeira:

\begin{equation}
\frac{d^{2}\xi \left( t\right) }{dt^{2}}+\Omega ^{2}\,\xi \left( t\right) =0,
\label{amo2}
\end{equation}%
com%
\begin{equation}
\Omega =\sqrt{\omega _{0}^{2}-\gamma ^{2}}.
\end{equation}%
Para o caso subamortecido ($\gamma <\omega _{0}$), a equa\c{c}\~{a}o (\ref%
{amo2}) \'{e} reconhecida como a equa\c{c}\~{a}o do oscilador harm\^{o}nico
simples e assim, tirando proveito do resultado constante na se\c{c}\~{a}o
anterior, podemos escrever solu\c{c}\~{a}o como%
\begin{equation}
x\left( t\right) =e^{-\gamma t}\left( A\cos \Omega t+B\mathrm{\,sen\,}\Omega
t\right) .
\end{equation}%
Para o caso criticamente amortecido ($\gamma =\omega _{0}$), a equa\c{c}\~{a}%
o (\ref{amo2}) torna-se desprovida de embara\c{c}os e exibe a solu\c{c}\~{a}%
o linear. Da\'{\i},
\begin{equation}
x\left( t\right) =e^{-\gamma t}\left( A+Bt\right) .
\end{equation}%
Finalmente, para o caso superamortecido ($\gamma >\omega _{0}$) a situa\c{c}%
\~{a}o torna-se constrangedora porque $\Omega $ manifesta-se como um n\'{u}%
mero imagin\'{a}rio, inviabilizando assim a busca de solu\c{c}\~{o}es por
interm\'{e}dio de s\'{e}ries de Fourier. Mais uma vez podemos tirar proveito
da solu\c{c}\~{a}o do oscilador harm\^{o}nico simples, fazendo uso das
identidades tri\-go\-no\-m\'{e}\-tri\-cas $\mathrm{\,sen\,}i\theta =i\mathrm{%
\,senh\,}\theta $ e $\cos i\theta =$ cosh$\,\theta $ para escrever%
\begin{equation}
\xi \left( t\right) =a\cosh \widetilde{\gamma }t+b\mathrm{\,senh\,}%
\widetilde{\gamma }t,
\end{equation}%
onde%
\begin{equation}
\widetilde{\gamma }=\sqrt{\gamma ^{2}-\omega _{0}^{2}}.
\end{equation}%
Escrevendo as fun\c{c}\~{o}es hiperb\'{o}licas em termos de exponenciais
temos que a solu\c{c}\~{a}o do caso superamortecido pode ser posta na forma%
\begin{equation}
x\left( t\right) =e^{-\gamma t}\left( Ae^{\widetilde{\gamma }t}+Be^{-%
\widetilde{\gamma }t}\right) .
\end{equation}%
\'{E} instrutivo observar que esta solu\c{c}\~{a}o tem o comportamento assint%
\'{o}tico prescrito anteriormente haja vista que $\widetilde{\gamma }<\gamma
$.

\section{Conclus\~{a}o}

Em suma, apresentamos as s\'{e}ries de Fourier como uma ferramenta
alternativa para a busca de solu\c{c}\~{o}es do oscilador harm\^{o}nico. A an%
\'{a}lise tem fulcro apenas no uso de s\'{e}ries de Fourier no caso do
oscilador harm\^{o}nico simples, acrescido da t\'{e}cnica de elimina\c{c}%
\~{a}o do termo de derivada primeira de equa\c{c}\~{o}es diferenciais de
segunda ordem no caso do oscilador harm\^{o}nico amortecido. Aos leitores
com disposi\c{c}\~{a}o de esp\'{\i}rito fica a tarefa de encontrar outros
sistemas f\'{\i}sicos cujas equa\c{c}\~{o}es diferenciais homog\^{e}neas
possam ser resolvidas com certa simplicidade pelo m\'{e}todo do
desenvolvimento da solu\c{c}\~{a}o em s\'{e}ries de Fourier.

\bigskip

\noindent{\textbf{Agradecimentos}}

O autor \'{e} grato ao CNPq pelo apoio financeiro.

\end{document}